\documentclass[twocolumn,nofootinbib,aps,superscriptaddress,preprintnumbers]{revtex4}

\usepackage{epsfig}
\usepackage{graphicx}
\usepackage[hypertex]{hyperref}

\newcommand{\nn}{\nonumber}
\newcommand{\beq}{\begin{equation}}
\newcommand{\eeq}{\end{equation}}
\newcommand{\beqa}{\begin{eqnarray}}
\newcommand{\eeqa}{\end{eqnarray}}

\def\d{{\rm d}}
\def\lqcd{\Lambda_{\rm QCD}}
\newcommand{\mXcut}{\ensuremath{m_X^{\rm cut}}}
\newcommand{\pxp}{\ensuremath{p_X^+}}
\newcommand{\pxph}{\ensuremath{\hat p_X^+}}

\newcommand{\GeV}{{\rm GeV}}
\newcommand{\ellpm}{\ell^+\ell^-}

\def\OMIT#1{}

\begin{document}

\preprint{\vbox{\hbox{MIT--CTP 3707} \hbox{LBNL--59134} \hbox{hep-ph/0512191}}}

\title{\boldmath Universality and $m_X$ cut effects in $B\to X_s \ell^+ \ell^-$}

\author{Keith S.M.\ Lee}
\affiliation{Center for Theoretical Physics, Massachusetts Institute of
Technology, Cambridge, MA 02139}

\author{Zoltan Ligeti}
\affiliation{Ernest Orlando Lawrence Berkeley National Laboratory,
University of California, Berkeley, CA 94720}
\affiliation{Center for Theoretical Physics, Massachusetts Institute of
Technology, Cambridge, MA 02139}

\author{Iain W.\ Stewart}
\affiliation{Center for Theoretical Physics, Massachusetts Institute of
Technology, Cambridge, MA 02139}

\author{Frank J.\ Tackmann}
\affiliation{Ernest Orlando Lawrence Berkeley National Laboratory,
University of California, Berkeley, CA 94720}

\begin{abstract}

The most precise comparison between theory and experiment for the $B\to X_s
\ell^+\ell^-$ rate is in the low $q^2$ region, but the hadronic uncertainties
associated with an experimentally required cut on $m_X$ potentially spoil the
search for new physics in these decays.  We show that a $10$--$30\%$ reduction
of $\d\Gamma(B\to X_s \ell^+\ell^-)/\d q^2$ due to the $m_X$ cut  can be
accurately computed using the $B\to X_s\gamma$ shape function. The effect is
universal for all short distance contributions in the limit $m_X^2 \ll m_B^2$,
and this universality is spoiled neither by realistic values of the $m_X$ cut
nor by $\alpha_s$ corrections. Both the differential decay rate and
forward-backward asymmetry with an $m_X$ cut are computed.

\end{abstract}

\maketitle

\section{Introduction}

In the standard model (SM) the flavor-changing neutral current process $B\to X_s
\ell^+\ell^-$ does not occur at tree level and is thus a sensitive probe of new
physics.  Predicting its rate involves integrating out the $W$, $Z$, and $t$ at
a scale of order $m_W$ by matching on to the
Hamiltonian~\cite{Grinstein:1988me,Buchalla:1995vs}
\beq\label{HW}
 H_W = -\frac{G_F}{\sqrt2}\, V_{tb} V^*_{ts}
  \bigg[ \sum_{i=1}^{6} C_i\, {O}_i + \frac{1}{4\pi^2}
\sum_{i=7}^{10} C_i\, O_i \bigg] ,
\eeq 
evolving to $\mu=m_b$, and computing matrix elements of $H_W$.  Here $O_1-O_6$
are four-quark operators and
\beqa
O_7 &=& \overline{m}_b  
   \, \bar s \sigma_{\mu\nu} e F^{\mu\nu} P_R b , \nn \\
O_8 &=&  \overline{m}_b  
  \, \bar s \sigma_{\mu\nu} gG^{\mu\nu} P_R b , \nn \\
O_9 &=& e^2 
  (\bar s \gamma_\mu P_L b) (\bar\ell \gamma^\mu \ell) , \nn \\
O_{10} &=& e^2 
  (\bar s \gamma_\mu P_L b) (\bar\ell \gamma^\mu\gamma_5 \ell) ,
\eeqa
where $P_{L,R} = (1\mp\gamma_5)/2$. Measurements of $C_{7,8,9,10}$ probe
flavor-changing neutral currents and test the SM. This can be done with the
dilepton invariant mass spectrum, $d\Gamma/dq^2$, with $q^2 = (p_{\ell^+} +
p_{\ell^-})^2$. It is calculable in an operator product expansion (OPE), and the
nonperturbative corrections are ${\cal
  O}(\lqcd^2/m_b^2)$~\cite{Falk:1993dh,Ali:1996bm}.  The matching and anomalous
dimensions for $C_i$ are known at next-to-next-to-leading log (NNLL) order, as
are the perturbative QCD corrections to the matrix elements of
$O_i$~\cite{nnll,Ghinculov:2002pe,Ghinculov:2003qd} (except small $O_{3-6}$
terms).

A complication in $B\to X_s\ell^+\ell^-$ compared with $B\to X_s\gamma$ is that
the long distance contributions, $B\to J/\psi X_s$ and $\psi' X_s$ followed by
$J/\psi,\, \psi' \to \ell^+\ell^-$, are two orders of magnitude above the
short distance prediction, a fact which is not well understood. Therefore,
either theory and data are both interpolated, or the short distance calculation
is compared with the data for $q^2 < m_{J/\psi}^2$ or $q^2 > m_{\psi'}^2$.  The
low $q^2$ region, $q^2 < 6\,\GeV^2$, allows the most precise comparison with the
SM, but requires a cut on the invariant mass of the hadronic final state, $m_X <
\mXcut$.  In the latest Belle analysis $\mXcut = 2\,\GeV$~\cite{Iwasaki:2005sy},
while Babar uses $\mXcut = 1.8\,\GeV$~\cite{Aubert:2004it}.  This cut is to
remove backgrounds and will likely be required for quite some
time~\cite{Berryhill}.  So far, its effect has been studied only in the
Fermi-motion model~\cite{Ali:1998ku}.  [The high $q^2$ region is unaffected by
the $m_X$ cut, but the rate is lower, and calculating it involves an expansion
in $\lqcd/(m_b-\sqrt{q^2})$.]

In this letter we compute the $B\to X_s\ell^+\ell^-$ rate with an $m_X$ cut in
the low $q^2$ region in a model-independent framework.  For $(\mXcut)^2 = {\cal
  O}(\lqcd m_b)$, the local OPE used in all earlier analyses breaks down and
must be replaced by an OPE involving $b$ quark distribution functions (shape
functions), as explained below. We will compute
\beq\label{epsdef}
 \Gamma_{ij}^{\rm cut} = \int_{q_1^2}^{q_2^2}\! \d q^2 \int_0^{\mXcut}\!\!
  \d m_X\, {\rm Re}(c_i c_j^*)\, \frac{\d^2\Gamma_{ij}}{\d q^2 \d m_X} \,,
\eeq
and study the ratios
\beq\label{eta}
  \eta_{ij}\big(\mXcut, q_1^2, q_2^2\big) = \frac{\Gamma_{ij}^{\rm
      cut}}{\Gamma_{ij}^{0}} \,.
\eeq
For convenience we define normalization factors
\beqa\label{gam0}
  \Gamma_{ij}^0  &=&   \frac{\Gamma_0}{m_B^5}
  \int_{q_1^2}^{q_2^2}\! \d q^2\, {\rm Re}(c_i c_j^*) 
  {(m_b^2-q^2)^2\over m_b^3}\, G_{ij} \,, \nn\\
  \Gamma_0 &=& \frac{G_F^2 m_B^5}{192\pi^3}\, \frac{\alpha^2_{\rm em}}{4\pi^2}\, 
|V_{tb} V_{ts}^*|^2 \,,
\eeqa
with kinematic dependence $G_{99}=G_{00} = (2q^2+m_b^2)$, $G_{77} =
4m_B^2(1+2m_b^2/q^2)$, and $G_{79} = 12m_Bm_b$.  Here and
below, $m_b$ is a short distance mass, such as $m_b^{1S}$~\cite{Hoang:1998hm}.  
In Eqs.~(\ref{epsdef}--\ref{gam0}), $ij = \{77,\, 99,\, 00,\, 79\}$ label
contributions of time-ordered products of operators, $T\{O_j^\dagger,
O_i\}$. The total  decay rate with cuts is the sum of  these contributions,
\beq\label{totalrate}
\Gamma^{\rm cut} = \Gamma\, \bigg|_{ \stackrel{\scriptstyle m_X<m_X^{\rm
  cut}}{q_1^2<q^2<q_2^2} } = \sum_{ij} \Gamma_{ij}^{\rm cut} \,.
\eeq
We will also study $\eta_{ij}' = \eta_{ij}'(p_X^{+\rm cut},q_1^2,q_2^2)$, which
differ from $\eta_{ij}$ by the replacement of $m_X$ by $p_X^+ = E_X-|\vec p_X|$:
\beq\label{eps2def}
  \eta_{ij}' = \frac{1}{\Gamma_{ij}^0} 
    \int_{q_1^2}^{q_2^2}\! \d q^2 \int_0^{p_X^{+\rm cut}}\!\!\!\!\!\!
  \d p_X^+\, {\rm Re}(c_i c_j^*)\, \frac{\d^2\Gamma_{ij}}{\d q^2 \d p_X^+} \,.
\eeq

In Eqs.~(\ref{epsdef}--\ref{eps2def}) the short distance coefficients
$c_{7,9,0}$ track the $C_{7,9,10}$ dependence in Eq.~(\ref{HW}) that one would
like to measure.  Here $c_7 = C_7^{\rm mix}(q^2)$, $c_9 = C_9^{\rm mix}(q^2)$,
and $c_0 = C_{10}$ can be obtained from local OPE
calculations~\cite{Buras:1994dj} at each order, as discussed in
Ref.~\cite{Lee:2005pk}. 

The $\eta_{ij}$'s contain the effects of the $m_X$ cut, and are defined with a
normalization that makes them less sensitive to the choice of $ij$.  At leading
order in $\lqcd/m_b$ and $\alpha_s$, $\eta_{ij}$ give the fraction of events
with $m_X < \mXcut$, and $\eta_{ij} = 1$ for $\mXcut = m_B$.  This
interpretation is altered at subleading order by $\alpha_s$ corrections, but
knowing $\eta_{ij}$ at a given order in perturbation theory is still sufficient
to determine $\Gamma_{ij}^{\rm cut}$ and thus the total rate with cuts in
Eq.~(\ref{totalrate}), at this order.  In principle, $\eta_{ij}$ depend in a
nontrivial way on $ij$ (and $q_1^2$ and $q_2^2$) due to different dependence on
kinematic variables, $\alpha_s$ corrections, etc.  At leading order in
$\lqcd/m_b$, we demonstrate that $\eta_{ij}$ are actually independent of the
choice of $ij$, a property which we call ``universality".  We first show this
formally in section~II at leading order in $p_X^+ /m_B \ll 1$ for
$\eta'(p_X^{+{\rm cut}})$. Then in section~III we demonstrate it numerically for
the experimentally relevant $\eta(\mXcut)$, including the $\alpha_s$ corrections
and phase space effects.

We also compute the decay rate for current experimental cuts. We find that the
rate is sensitive to the choice of the cut, and that the cut causes a reduction
in the rate by $10$--$30\%$.  Since the same shape function occurs in $B\to X_s
\ell^+\ell^-$, $X_u\ell\bar\nu$, and $X_s\gamma$, the \mXcut\ or $p_X^{+\rm
cut}$ dependence in one can be accurately determined from the others, and the
magnitude of the reduction can be computed quite accurately.  Alternatively,
instead of using the theoretical computation of the $\mXcut$ dependence,
universality can be exploited to remove the main uncertainties, by normalizing
the $B\to X_s\ell^+\ell^-$ rate to $B\to X_u\ell\bar\nu$.

\section{\boldmath $m_X$ cut effects at leading order}

\begin{figure}[t]
\centerline{\includegraphics[width=0.9\columnwidth]{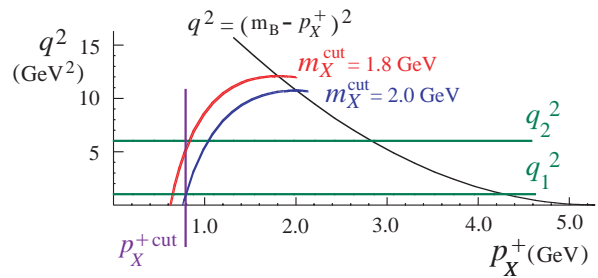}}
\vskip -0.2cm
\caption{Phase space cuts.  A substantial part of the rate for $q_1^2< q^2<
q_2^2$ falls in the rectangle bounded by $p_X^+ < p_X^{+{\rm cut}}$.}
\label{fig_kinematics}
\end{figure}

For simplicity, consider the kinematics in the $B$ meson's rest frame.  
Since $q = p_B - p_X$,
\beq
2m_B E_X = m_B^2 + m_X^2 - q^2 .
\eeq
If $m_X^2 \ll m_B^2$ and $q^2$ is not near $m_B^2$, then $E_X = {\cal O}(m_B)$.
Since $E_X^2 \gg m_X^2$, $p_X$ is near the light-cone, with $p_X^+ = E_X - |\vec
p_X\!| = {\cal O}(\lqcd)$ and $p_X^- = E_X + |\vec p_X\!| = {\cal O}(m_B)$.  Of
the variables symmetric in $p_{\ell^+}$ and $p_{\ell^-}$ ($p_X^\pm$,
$E_X$, $q^2$, $m_X^2$), only two are independent, and we work with $q^2 $ and 
$p_X^+$ or $m_X$. The phase space cuts are shown in~Fig.~\ref{fig_kinematics}.

For the $p_X^+ \ll p_X^-$ region, factorization of the form $\d\Gamma = H J
\otimes \hat f^{(0)}$ has been proven for semileptonic and radiative $B$
decays~\cite{Korchemsky:1994jb}, where $H$ contains perturbative physics at
$\mu_b \sim m_b$, $J$ at $\mu_i \sim \sqrt{\lqcd m_b}$, and $\hat
f^{(0)}(\omega)$ is a universal nonperturbative shape
function~\cite{Neubert:1993ch}. This factorization also applies for $B\to
X_s\ell^+\ell^-$ with the same $\hat f^{(0)}$, as long as $q^2$ is not
parametrically small~\cite{Lee:2005pk}. 

In the $q^2< 6\,{\rm GeV}^2$ region, $|C_9^{\rm mix}(q^2,\mu_0=4.8\,\GeV)|=4.52$
to better than $1\%$, and can be taken to be constant.  We neglect $\alpha_s$
corrections in this section and find
\beqa\label{leading}
\frac{\d\Gamma}{\d p_X^+ \d q^2}
&=&  \hat f^{(0)}(\pxp)\, \frac{\Gamma_0}{m_B^5}\,
  \frac{[(m_B - \pxp)^2 - q^2]^2}{(m_B - \pxp)^3} \nn\\
&\times& \Big\{ ( |C_9^{\rm mix}|^2 + C_{10}^2 )
  \big[ 2 q^2 + (m_B - \pxp)^2 \big] \nn\\
&&{} + 4m_B^2\, | C_7^{\rm mix}|^2 \Big[ 1 + \frac{2(m_B - \pxp)^2}{q^2} \Big] 
  \nn\\
&&{} + 12m_B\, {\rm Re}\big[ C_7^{\rm mix} C_9^{\rm mix}{}^*\big]
  (m_B - \pxp) \Big\} , \qquad
\eeqa
where $\hat f^{(0)}(\omega)$ has support in $\omega\in [0,\infty)$. As a
function of $p_X^+$, the kinematic terms in Eq.~(\ref{leading}) vary {\em only}
on a scale $m_B$, while $\hat f^{(0)}(\pxp)$ varies on a scale $\lqcd$.  Writing
$m_B = m_b + \bar\Lambda$ and expanding in $(\pxp-\bar\Lambda)/m_B$ decouple the
$p_X^+$ and $q^2$ dependences in Eq.~(\ref{leading}), and give exactly the local OPE
prefactors, $(m_b^2-q^2)^2\, G_{ij}(q^2)$, used in Eq.~(\ref{gam0}).  For
$\eta_{ij}'(p_X^{+{\rm cut}},q_1^2,q_2^2)$, the \pxp\ integration is over a
rectangle in Fig.~\ref{fig_kinematics}, whose boundaries do not couple $p_X^+$
and $q^2$. Thus, with the above expansion, we find $\eta_{ij}' =\eta'$, where
\beq
  \eta'= \int\! \d p_X^+\, \hat f^{(0)}(\pxp) \,,
\eeq
independent of
$ij$ and $q_{1}^2$, $q_2^2$.  While the $m_X$ cut retains more events than the $\pxp$
cut, the latter may give theoretically cleaner constraints on short distance
physics when statistical errors become small.

\begin{figure}[t]
\centerline{\includegraphics[width=0.9\columnwidth]{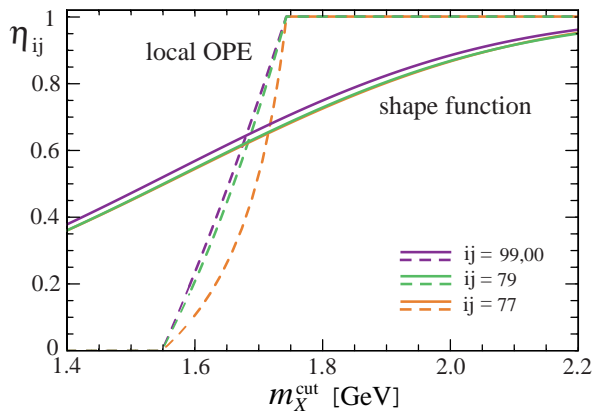}}
\caption{$\eta_{ij}(\mXcut, 1\,\GeV^2, 6\,\GeV^2)$ as functions of
$\mXcut$.  The dashed curves show the local OPE result, the solid curves include
the leading shape function effects.  The uppermost, middle, and lowest curves
are  $\eta_{00,99}$, $\eta_{79}$, and $\eta_{77}$, respectively.}
\label{fig_eps0}
\end{figure}

The effect of the $m_X$ cut is $q^2$ dependent, because the upper limit of the
$p_X^+$ integration is $q^2$ dependent, as shown in Fig.~\ref{fig_kinematics}. 
When we include the full $p_X^+$ dependence in Eq.~(\ref{leading}), the universality
of $\eta_{ij}(m_X^{\rm cut}, q_1^2, q_2^2)$ is maintained to better than 3\%
for  $1\,\GeV^2 \leq q_1^2 \leq 2\,\GeV^2$, $5\,\GeV^2 \leq q_2^2 \leq
7\,\GeV^2$, and $\mXcut \geq 1.7\,\GeV$, because the region where the $p_X^+$
and $q^2$ integration limits are coupled has a small effect on the $ij$
dependence.  This is exhibited in Fig.~\ref{fig_eps0}, where the solid curves
show $\eta_{ij}(\mXcut, 1\,\GeV^2, 6\,\GeV^2)$ with the shape function set to
model~1 of~\cite{Tackmann:2005ub}, with $m_b^{1S}=4.68\,\GeV$ and $\lambda_1$
from~\cite{Bauer:2004ve}.  (Taking $q_1^2 = 1\,\GeV^2$ instead of $4m_\ell^2$
increases the sensitivity to $C_{9,10}$, but one may be concerned by local
duality\,/\,resonances near $q^2 = 1\,\GeV^2$.  To estimate this uncertainty,
assume the $\phi$ is just below the cut and ${\cal B}(B\to X_s\phi) \approx 10
\times {\cal B}(B\to K^{(*)}\phi)$.  Then $B\to X_s\phi\to X_s\ell^+\ell^-$ is
$\sim\!2\%$ of the $X_s\ell^+\ell^-$ rate.)

The local OPE results for $\eta_{ij}(\mXcut,q_1^2,q_2^2)$ are obtained by
replacing $\hat f^{(0)}(\pxp)$ by $\delta(\bar\Lambda-\pxp)$ in
Eq.~(\ref{leading}).  Performing the $p_X^+$ integral sets $(m_B-\pxp)=m_b$ and
implies $m_X^2 > \bar\Lambda(m_B-q^2/m_b)$.  This makes the lower limit on $q^2$
equal $\max\{q_1^2,\, m_b[m_B-(\mXcut)^2/\bar\Lambda]\}$, and so the
$\eta_{ij}$'s depend on the shape of $\d\Gamma_{ij}$.  In Fig.~\ref{fig_eps0}
the local OPE results are shown by dashed lines, and clearly $\eta_{77}\ne
\eta_{99}$.  However, the local OPE is not applicable for $p_X^+\sim \lqcd$.

The universality of $\eta_{ij}$ found here could be broken by $\alpha_s$
corrections in the $H$ or $J$ functions, or by renormalization group
evolution, since these effects couple $p_X^+$ and $q^2$ and have been neglected
so far.  We consider these next.

\section{\boldmath Calculation and results at ${\cal O}(\alpha_s)$}

A complication in calculating $B\to X_s\ell^+\ell^-$ compared with $B\to
X_u\ell\bar\nu$ is that, in the evolution of the effective Hamiltonian down to
$m_b$, $C_9(\mu)$ receives a $\ln(m_W^2/m_b^2)$ enhanced contribution from the
mixing of $O_2$. Thus, formally, $C_9\sim {\cal O}(1/\alpha_s)$, and
conventionally one expands the amplitude in $\alpha_s$, treating $\alpha_s
\ln(m_W^2/m_b^2) = {\cal O}(1)$~\cite{Buras:1994dj}.  In the local OPE this is
reasonable, since the nonperturbative corrections are small, and at
next-to-leading log (NLL) all dominant terms in the rate are included.  However,
in the shape function region nonperturbative effects are ${\cal O}(1)$ and only
the rate is calculable, not the amplitude.  With the traditional counting, the
$C_9^2$ contribution to the rate would be needed to ${\cal O}(\alpha_s^2)$
before the $C_{10}^2$ terms could be included.

This would be a bad way to organize the perturbative corrections (numerically
$|C_9(m_b)| \approx |C_{10}|$).  It can be circumvented by using a ``split
matching" procedure to decouple the perturbation series above and below the
scale $m_b$~\cite{Lee:2005pk}.  This allows us to consider the short distance
coefficients $C_7^{\rm mix}$, $C_9^{\rm mix}$, and $C_{10}$ as ${\cal O}(1)$
numbers when organizing the perturbation theory at $m_b^2$ and $m_b\lqcd$.

The rate and the forward-backward asymmetry are
\beqa\label{rates}
\frac{\d^2\Gamma}{\d q^2 \d p_X^+}
  &=& \frac{\Gamma_0}{m_B^2}\,
  H(q^2,  p_X^+)\,  F^{(0)}(p_X^+,p^-)\,, \nn\\[4pt]
\frac{\d^2 A_{\rm FB}}{\d q^2 \d p_X^+}
  &=& \frac{\Gamma_0}{m_B^2}\,
  K(q^2, p_X^+)\,  F^{(0)}(p_X^+,p^-) \,,
\eeqa
where $p^- = m_b - q^2/(m_B-p_X^+)$.  The hard functions $H$ and $K$ were
computed in~\cite{Lee:2005pk} using soft-collinear effective theory
(SCET)~\cite{sceta,scetb} and split matching.  This factorizes the dependence on
scales above and below $m_b$ as $\Gamma_{ij} \sim H_1(\mu_0) H_2(\mu_b)
F^{(0)}(\mu_b)$, with separate $\mu_0$ and $\mu_b$ independence.  Up to the
order one is working at, $H_1$ is $\mu_0$ independent, the $\mu_b$ dependence in
$H_2$ and $F^{(0)}$ cancels, and $F^{(0)}$ is $\mu_i$ independent.  The shape
function model is specified at $\mu_\Lambda$.  The convolution of jet and shape
functions at NLL including $\alpha_s$ corrections is
\begin{widetext}
\beqa
F^{(0)}(p^+_X, p^-) 
&=& U_H(p^-,\mu_i,\mu_b)\, \bigg\{ \hat f^{(0)}\big( p^+_X,\mu_i \big)
  + \frac{\alpha_s(\mu_i)C_F}{4\pi}\, \bigg[ \Big( 
  2 \ln^2\frac{p^+_X p^-}{\mu_i^2} -  3 \ln \frac{p^+_X p^-}{\mu_i^2} 
  + 7- \pi^2 \Big)\, \hat f^{(0)}\big( p^+_X,\mu_i \big) \nn\\
&& + \int_0^1 \frac{\d z}{z}\, \Big( 4 \ln\frac{z p^+_X p^-}{\mu_i^2} - 3 \Big)
  \Big( \hat f^{(0)}\big(p_X^+ (1-z),\mu_i\big)
   - \hat f^{(0)}\big(p_X^+,\mu_i \big)\Big) \bigg]\bigg\} ,\nn\\[6pt]
\hat f^{(0)}(\omega,\mu_i) &=& \frac{e^{V_S(\mu_i,\mu_\Lambda)}}{\Gamma(1+\eta)}
  \left(\frac{\omega}{\mu_\Lambda}\right)^\eta
  \int_0^1 \d t\, \hat f^{(0)} \big[\omega(1 - t^{1/\eta}), \mu_\Lambda\big],
\eeqa
where $\eta = (16/25) \ln[\alpha_s(\mu_\Lambda)/\alpha_s(\mu_i)]$, $U_H$ was
computed in Ref.~\cite{sceta}, the one-loop jet function in
Ref.~\cite{bm04,blnp04}, and the shape function evolution up to $\mu_i$ in
Refs.~\cite{sceta,blnp04} (for earlier calculations, see
Refs.~\cite{Korchemsky:1994jb,Leibovich:2000ig}).  The hard coefficients $H$ and
$K$ for $B\to X_s\ell^+\ell^-$ are
\beqa\label{HK}
H( q^2, p_X^+) &=&
  \frac{[(1- \pxph)^2- \hat q^2]^2}{(1- \pxph)^3}\,
  \bigg\{ \big[ |C_9^{\rm mix}(s,\mu_0)|^2 + C_{10}^2 \big] 
  \Big[ 2\hat q^2\, \Omega_A^2(s,\mu_b) + (1 - \pxph)^2\,
  \Omega_B^2(s,\pxph,\mu_b) \Big] \nn\\*
&&{} + 4 | C_7^{\rm mix}(\mu_0)|^2 \Big[ \Omega_C^2(s,\mu_b)
  + \frac{2(1 - \pxph)^2}{\hat q^2}\, \Omega_D^2(s,\mu_b) \Big]
  + 12 {\rm Re}\big[ C_7^{\rm mix}(\mu_0) C_9^{\rm mix}(s,\mu_0)^*\big]
  (1 - \pxph) \Omega_E(s,\mu_b) \bigg\} \nn\\[4pt]
K(q^2, p_X^+) &=& - \frac{3 \hat q^2 [(1 - \pxph)^2- \hat q^2]^2 }
  {(1 - \pxph)^3}\, \Omega_A(s,\mu_b)\, {\rm Re} \Big\{ C_{10}^* 
  \Big[ C_9^{\rm mix}(s,\mu_0) \Omega_A(s,\mu_b)
  + \frac{2(1- \pxph )}{\hat q^2}\,
  C_7^{\rm mix}(\mu_0) \Omega_D(s,\mu_b) \Big] \Big\},
\eeqa
where $s = q^2/m_b^2$, $\hat q^2 = q^2/m_B^2$, $\pxph = \pxp/m_B$, and
\beqa
\Omega_A &=& 1 + \frac{\alpha_s}{\pi}\, \omega_a^V\!(s,\mu_b) \,, \qquad\,
\Omega_B = 1 + \frac{\alpha_s}{\pi} \Big[ \omega_a^V\!(s,\mu_b) + 
  \frac{(1-\pxph)^2 - \hat q^2}{2(1-\pxph)^2}\,
  \omega_b^V\!(s) + \omega_c^V\!(s) \Big] , \nn\\ 
\Omega_C &=& 1 + \frac{\alpha_s}{\pi}\, \omega_a^T(s,\mu_b) \,, \qquad
\Omega_D = 1 + \frac{\alpha_s}{\pi} \big[ \omega_a^T(s,\mu_b) 
  - \omega_c^T(s) \big] \,, \qquad
\Omega_E = \frac13 \big( 2 \Omega_A\Omega_D + \Omega_B \Omega_C\big) \,.
\eeqa
\end{widetext}
Here $\alpha_s = \alpha_s(\mu_b)$ and $\omega_i^{V,T}$ are defined in
Ref.~\cite{Lee:2005pk}.

For each shape function model, the deviations of the $\eta_{ij}$'s from being
universal, with all NLL corrections, are still below 3\%. Thus, the picture
of universality in Fig.~\ref{fig_eps0} remains valid at NLL order. For this
reason we can explore the overall shift by just studying $\eta_{00}$.

\begin{figure}[b!]
\centerline{\includegraphics[width=0.9\columnwidth]{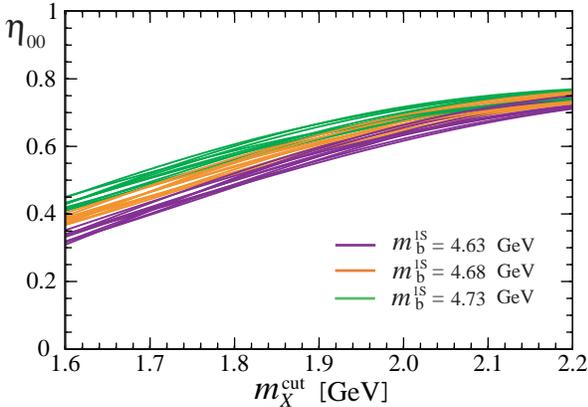}}
\caption{$\eta_{00}(\mXcut, 1\,\GeV^2, 6\,\GeV^2)$ as a function of $\mXcut$. 
The orange, green and purple (medium, light, dark) curves show $m_b^{1S} =
4.68\,\GeV$, $4.63\,\GeV$, and $4.73\,\GeV$, respectively.}
\label{fig:3}
\end{figure}
In Fig.~\ref{fig:3} we plot $\eta_{00}(\mXcut, 1\,\GeV^2, 6\,\GeV^2)$, including
the $\alpha_s$ corrections.  We use ten different models for $\hat f^{(0)}$. Our
base model has five parameters, three of which are chosen to obey the known
constraints on its moments~\cite{blnp04}, converted to the $1S$ mass scheme used
here. For each of five different choices of the remaining two parameters, we
choose two values of the scale, $\mu_\Lambda$, where the model is specified. The
choice of these ten models is guided by making them consistent with the $B\to
X_s\gamma$ data.  The ten orange, green and purple (medium, light, dark) curves
correspond to $m_b^{1S} = 4.68\,\GeV$, $4.63\,\GeV$, and $4.73\,\GeV$,
respectively, with the central values $\mu_0 = \mu_b = 4.8\,\GeV$ and $\mu_i =
2.5\,\GeV$.  The curves with slightly lower [higher] values of $\eta_{00}$ at
large \mXcut\ correspond to $\mu_\Lambda = 1.5\,\GeV$ [$2\,\GeV$]. The spread in
the curves gives our determination of the uncertainty from the choice of shape
function model and from $m_b$. For $\mXcut = 2\,\GeV$, varying $\mu_b$ in the
range $3.5\,\GeV < \mu_b < 7.5\,\GeV$ changes $\eta_{00}$ by $\pm6\%$.  We find
a $\pm5\%$ variation for $2\,\GeV < \mu_i < 3\,\GeV$.

Using the $c_i$'s at NLL, for $1\,\GeV^2 < q^2 < 6\,\GeV^2$ we obtain cut
branching ratios 
\beq
 \Gamma^{\rm cut}\, \tau_B = \bigg\{ \begin{array}{c}
  (1.20 \pm 0.15) \times 10^{-6}  \quad [\mXcut =1.8\,\GeV] \,, \\
  (1.48 \pm 0.14) \times 10^{-6} \quad [\mXcut =2.0\,\GeV] \,,
   \end{array}
\eeq
where uncertainties are included from $m_b$, $\mu_b$, $\mu_i$, and $\hat
f^{(0)}$.  Changing $\mu_0$ to 3.5\,\GeV\ (10\,\GeV) changes both of these rates
by $-2$\% ($+7$\%), and this uncertainty will be reduced by including NNLL
corrections~\cite{nnll,Ghinculov:2002pe,Ghinculov:2003qd}.

The largest source of universality breaking in the $\eta_{ij}$'s and one of the
largest uncertainties in the cut rate is due to subleading shape functions, which
affect the rate by $\sim$\,5\% for $\mXcut = 2\,\GeV$ and by $\sim$\,10\% for
$\mXcut = 1.8\,\GeV$~\cite{next}.

If the \mXcut\ dependence were not universal, it would modify the zero of the
forward-backward asymmetry, $A_{\rm FB}(q_0^2)=0$.  For $m_X^{\rm cut} =
2\,\GeV$ we find at NLL $\Delta q_0^2 \approx -0.04\,\GeV^2$, much below the
higher order uncertainties~\cite{Ghinculov:2002pe,Ghinculov:2003qd}.  However,
we obtain $q_0^2 = 2.8\,\GeV^2$, lower than earlier
results~\cite{Ghinculov:2002pe}.
The reason is that in the SCET calculation of $A_{\rm FB}$, using $K$ in
Eq.~(\ref{HK}), the pole mass $m_b^{\rm pole}$ never occurs, only $m_B-\pxp$ and
$\overline{m}_b$ (at this order, $C_7^{\rm mix} = (\overline{m}_b/m_B) C_7^{\rm
eff}$~\cite{Lee:2005pk}).  Thus, schematically, $q_0^2 \sim 2m_b
[\overline{m}_b(\mu_0) C_7^{\rm eff}(\mu_0)] / {\rm Re}[C_9^{\rm eff}(q_0^2)]$,
and there is no reason to expand $\overline{m}_b$ in terms of $m_b^{\rm pole}$.
 
In the above analysis, the nonperturbative shape function $f^{(0)}$ was
extracted from moments and the $B\to X_s\gamma$ energy spectrum, and this was
used as input in determining our $B\to X_s\ell^+\ell^-$ results. The overall
10\% theoretical uncertainty in this approach could be reduced by raising the
\mXcut.  An alternative approach would be to keep $m_X^{\rm cut}< m_D$ and
measure 
\beq 
R = {\Gamma^{\rm cut}(B\to X_s\ell^+\ell^-) \over 
  \Gamma^{\rm cut}(B\to X_u\ell\bar\nu)}\,, \eeq
with the same cuts used in the numerator and denominator. The dependence of the
semileptonic rate on $\mXcut$ is identical to that of $\Gamma_{00}^{\rm cut}$.
A measurement of $R$ bypasses the need for a shape function model, because we
found that the $m_X$-cut effects are universal to a very good approximation and
therefore cancel between the numerator and denominator of $R$.

In conclusion, we pointed out that the observed $B\to X_s\ell^+\ell^-$ rate in
the low $q^2$ region is sensitive to the experimental upper cut on $m_X$. The
reduction in the rate due to this cut is determined by the universal $B$ shape
function.  In the region of the experimental measurements an OPE exists only
for the decay rate and not for the amplitude, a fact that necessitates a reorganization of
the usual perturbation expansion.  Since one can use the shape function measured
in other processes, the sensitivity to new physics is not reduced.  We found
that the $\eta$'s for the different operators' contributions are universal to a
good approximation.  These results also apply for $B\to X_d\ellpm$, which may be
studied at a higher luminosity $B$ factory.  Subleading $\lqcd/m_b$ as well as
NNLL corrections to the rate and the forward-backward asymmetry will be studied
in a separate publication~\cite{next}.

\vspace{4pt} {\bf Acknowledgments~} We thank Bjorn Lange for helpful
discussions.  This work was supported in part by the Director, Office of
Science, and Offices of High Energy and Nuclear Physics of the U.S.\ Department
of Energy under the Contract DE-AC02-05CH11231 (Z.L.\ and F.T.), and the
cooperative research agreement DOE-FC02-94ER40818 (K.L.\ and I.S.). I.S. was
also supported in part by the DOE OJI program and by the Sloan Foundation.

\end{document}